%% file: note2308.tex
\documentclass[twocolumn,showpacs,aps,prd,superscriptaddress,floatfix,letter]{revtex4}
\usepackage{epsfig}
\usepackage{graphics}
\usepackage{graphicx}
\usepackage{amsmath}
\usepackage{rotating}
\usepackage{color}

\newcommand{\BaBarYear}       {11}
\newcommand{\BaBarNumber}     {006}
\newcommand{\SLACPubNumber} {14444}
 \newcommand{\BaBarType}      {PUB}

\input babarsym

\begin{document}

\begin{flushleft}
\babar-\BaBarType-\BaBarYear/\BaBarNumber \\
SLAC-PUB-\SLACPubNumber\\
\end{flushleft}

\begin{flushright}
\end{flushright}
\title{\boldmath Search for $b\to u$ Transitions 
in $\Bpm \to [\Kmp\pipm\piz]_D \Kpm$ Decays}
\input{authors_feb2011.tex}
\begin{abstract}\noindent

{\tolerance=0 We present a study of 
the decays $\Bpm \to D \Kpm$ with $D$ mesons
reconstructed in the $K^+\pi^-\pi^0$ or $K^-\pi^+\pi^0$ final states, 
where $D$ indicates a \Dz\ or a \Dzb\ meson. 
Using a sample of $474$ million \BB pairs collected with
the \babar\ detector at the PEP-II asymmetric-energy $e^+e^-$ collider
at SLAC, we measure the ratios $\textstyle R^{\pm} \equiv
\frac{\Gamma(\Bpm \to [\Kmp\pipm\pi^0]_D \Kpm)}{\Gamma(\Bpm \to
[\Kpm\pimp\pi^0]_D \Kpm)}$. We obtain $\rplus\ = \left(5^{+12}_{-10}\stat^{+2}_{-4}\syst\right)\times10^{-3}$ and
$\rminus\ = \left(12^{+12}_{-10}\stat^{+3}_{-5}\syst\right)\times10^{-3}$, from which we extract the upper limits
at 90\% probability: $\rplus<23\times10^{-3}$ and $\rminus<29\times10^{-3}$. Using these measurements, we
obtain an upper limit for the ratio \rB\ of the magnitudes of the $b \to u$ and $b \to c$
amplitudes $\rB<0.13$ at 90\% probability.}
\end{abstract}

\maketitle

\section{Introduction}
\begin{figure}[htb]
\begin{center}
\epsfig{figure=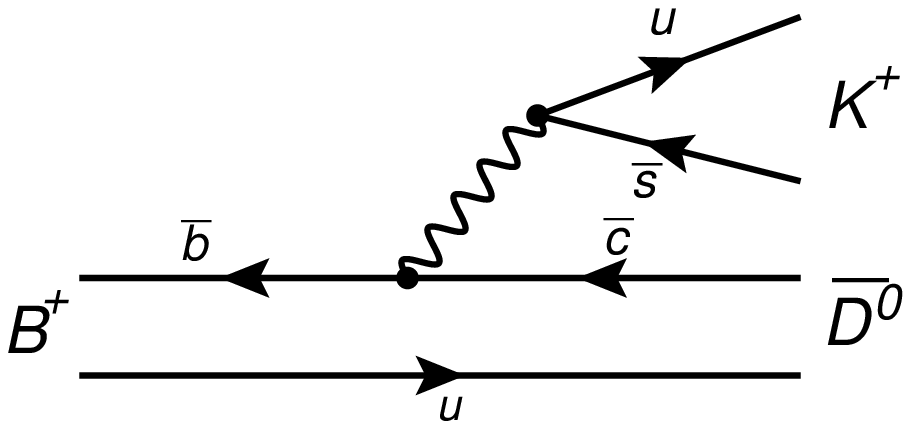,width=0.45\linewidth}\hskip 0.08\linewidth
\epsfig{figure=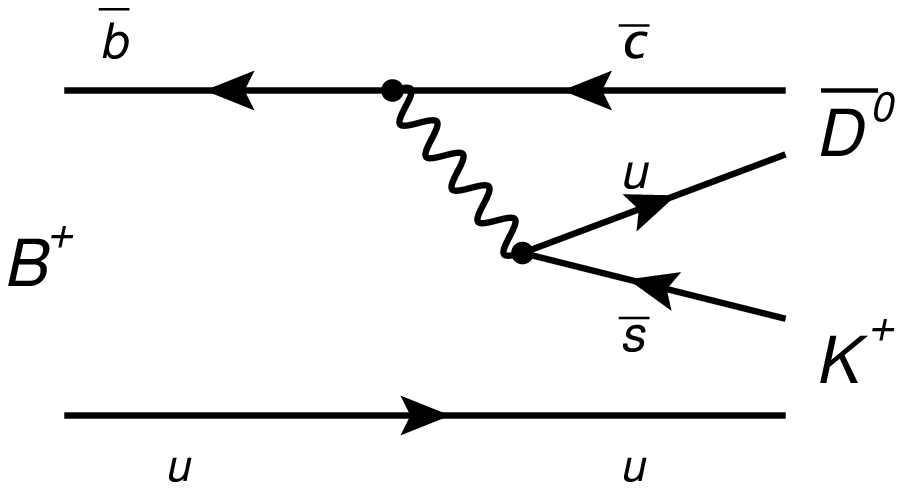,width=0.45\linewidth}\\\vskip 0.3cm
\epsfig{figure=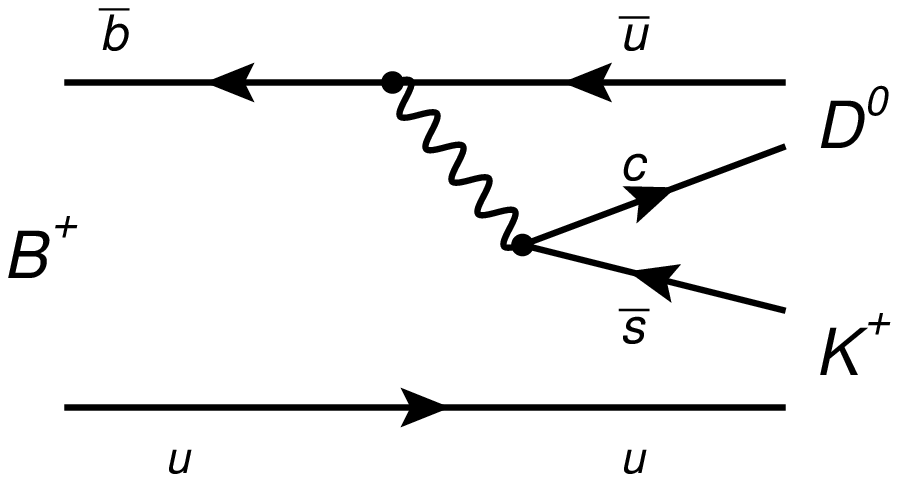,width=0.45\linewidth}\hskip 0.08\linewidth
\epsfig{figure=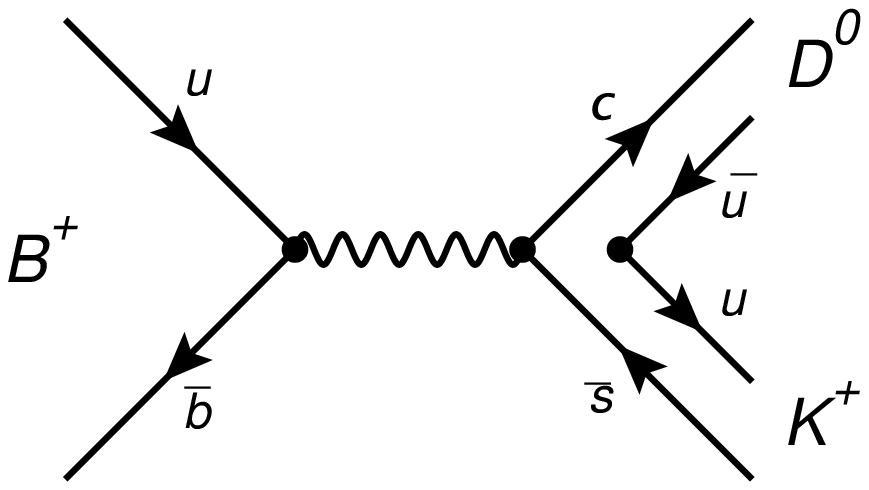,width=0.45\linewidth}
\end{center}
\caption{Feynman diagrams for $\Bp \to \Dzb \Kp$ (top, $\bbar \to \cbar$ transition) and $\Bp \rightarrow \Dz \Kp$ 
(bottom, $\bbar \to \ubar$ transition).\label{fig:feyn}}
\end{figure}

\CP violation effects are described in the Standard Model (SM) of
elementary particles with a single phase in the
Cabibbo-Kobayashi-Maskawa (CKM) quark mixing matrix
$V_{ij}$~\cite{cite:ckm}. One of the unitarity 
conditions for this matrix can be interpreted as 
a triangle in the plane of Wolfenstein parameters~\cite{cite:Wolf}, 
where one of the angles is 
$\gamma=\mbox{arg}\{-V^{*}_{ub}V_{ud}/V^{*}_{cb}V_{cd}\}$. Various
methods to determine \g\ using $\Bp\to D \Kp$ decays 
have been proposed~\cite{cite:GLW,cite:ADS,cite:DKDalitz}.
In this paper, we consider the decay channel $\Bp \rightarrow D
\Kp $ with $D \to \Km \pip \piz$~\cite{cite:charge} studied through the
Atwood-Dunietz-Soni (ADS) method~\cite{cite:ADS}. In this method the final 
state under consideration can be reached through $b \to c$ and $b \to u$
processes as indicated in Fig.~\ref{fig:feyn} that are followed by either 
Cabibbo-favored or Cabibbo suppressed \Dz decays. The interplay between different 
decay channels leads to a possibility to extract the angle \g alongside with other parameters
for these decays.

Following the ADS method, we search for $\Bp \to [\Km\pip\piz]_D \Kp$ events, where the 
favored $\Bp \to \Dzb \Kp$ decay, followed by the doubly-
Cabibbo-suppressed $\Dzb \to \Km\pip\piz$ decay, interferes with the
suppressed $\Bp \to \Dz \Kp$ decay, followed by the
Cabibbo-favored $\Dz \to \Km\pip\piz$ decay. These are called
``opposite-sign'' events because the two kaons in the final state
have opposite charges. We also reconstruct a larger sample of
``same-sign'' events, which mainly arise from the favored $\Bp
\to \Dzb \Kp$ decays followed by the Cabibbo-favored $\Dzb \to
\Kp\pim\piz$ decays. We define $f\equiv \Kp\pim\piz$ and $\bar
f\equiv \Km\pip\piz$. We extract
\begin{eqnarray}
\rplus&=\frac{\Gamma(\Bp\to [\bar f]_{D}\Kp)}{\Gamma(\Bp\to
[f]_{D}\Kp)},\label{eq:rmin}\\
\rminus&=\frac{\Gamma(\Bm\to [f]_{D}\Km)}{\Gamma\left(\Bm\to [\bar
f]_{D}\Km\right)}\label{eq:rplus}.
\end{eqnarray}
from the selected \Bp and \Bm samples, respectively. 

While our previous analysis~\cite{cite:ViolaADS} used another set of observables:
\begin{eqnarray}
\rads \equiv& \frac{\Gamma(\Bp \to [\bar f]_D \Kp) +\Gamma(\Bm \to
[f]_D \Km)} {\Gamma\left(\Bp \to [f]_D \Kp\right)
+\Gamma\left(\Bm \to [\bar f]_D \Km\right)},\\
\aads \equiv&   \frac{\Gamma\left(\Bm
\to [f]_D \Km\right)-\Gamma\left(\Bp \to [\bar f]_D \Kp\right)} {\Gamma\left(\Bp \to [\bar f]_D
\Kp\right)+\Gamma\left(\Bm \to [f]_D \Km\right)},\label{eq:rads}
\end{eqnarray}
we prefer to use observables defined in Eqs.~\ref{eq:rmin} and~\ref{eq:rplus} since their 
statistical uncertainties, which dominate in the final error 
of this measurement, are uncorrelated. 

The amplitude of the two-body \B decay can be written as
\begin{equation}
A(\Bp\to\Dz\Kp)=|A(\Bp\to\Dzb\Kp)|\rB e^{i\gamma} e^{i\deltaB},
\end{equation}
where $\textstyle \rB\equiv\frac{|A(\Bp\to \Dz\Kp)|}{|A(\Bp\to
\Dzb\Kp)|}$ is the ratio of the magnitudes of the $\b\to\u$ and $\b\to\c$ amplitudes,
\deltaB\ is the \CP conserving strong phase, and \g is the \CP\ violating weak phase.
For the three-body $D$ decay we use similarly defined variables:
\begin{eqnarray}
\rD^2 \equiv\frac{\Gamma\left(\Dz\to f\right)}{\Gamma\left(\Dz\to \bar f\right)}=
\frac{\int d\vec m\, A^2_{\rm DCS}(\vec m)}{\int d\vec m\, A^2_{\rm CF}(\vec m)},\\
\kD e^{i\deltaD}\equiv \frac{\int d\vec m\, A_{\rm DCS}(\vec m)A_{\rm CF}(\vec m)
e^{i\delta(\vec m)}}{\sqrt{\int d\vec m\, A_{\rm DCS}^2(\vec m)\int d\vec m\, A_{\rm
CF}^2(\vec m)}},
\end{eqnarray}
where $A_{\rm CF} (\vec m)$ and $A_{\rm DCS}(\vec m)$ are the magnitude of the 
Cabibbo-favored and doubly-Cabibbo-Suppressed amplitudes, respectively, $\delta(\vec m)$
is the relative strong phase, and $\vec m$ indicates a position in the $D$ Dalitz plot of 
squared invariant masses $[m^2_{K\pi}, m^2_{K\pi^0}]$. The parameter \kD, called the coherence factor, can
take values in the interval $[0,1]$.

Neglecting $D$-mixing effects, which in the SM
give negligible corrections to \g\ and do not affect the 
\rB\ measurement, the ratios \rplus\ and \rminus\ are related to
the $B$- and $D$-mesons' decay parameters through the following relations:
\begin{eqnarray}
\rplus&=\rB^2+\rD^2+2\rB\rD\kD\cos(\g+\delta),\label{eq:main1}\\
\rminus&=\rB^2+\rD^2+2\rB\rD\kD\cos(\g-\delta)\label{eq:main2},
\end{eqnarray}
with $\delta=\deltaB+\deltaD$. The values
of \kD\ and \deltaD\ measured by the CLEO-c
collaboration~\cite{cite:CLEO}, $\kD=0.84\pm 0.07$ and 
$\deltaD =(47^{+14}_{-17})^{\circ}$, are used in the 
signal yield estimation and
\rB\ extraction. The ratio \rD\ has been
measured in different experiments and we take the average value  
$\rD^2 = (2.2\pm 0.1)\times 10^{-3}$~\cite{cite:PDG}. Its value
is small compared to the present determination of
\rB, which is taken to be $(0.106\pm0.016)$~\cite{cite:UTFIT}. 
According to Eqs.~\ref{eq:main1} and~\ref{eq:main2}, this implies that the 
measurements of ratios $R^{\pm}$ is mainly sensitive to
\rB. For the same reason, the
sensitivity to \g is reduced, and
therefore the main aim of this analysis is to measure \rplus, \rminus, and \rB.
The current high precision on \rB\ is based on several earlier analyses 
by the \babar~\cite{cite:ViolaADS,cite:BaBarADS,cite:BaBarDalitz,cite:BaBarGLW}, 
BELLE~\cite{cite:BelleADS,cite:BelleDalitz,cite:BelleGLW}, and CDF~\cite{cite:CDFGLW} collaborations.

This paper is an update of our previous analysis~\cite{cite:ViolaADS} based on 
$226\times 10^{6}$ \BB pairs and resulting in a measurement of $\rads=(13^{+12}_{-10})\times 10^{-3}$, 
which was translated into 
the $95\%$ confidence level limit $\rB < 0.19$. 

The results 
presented in this paper are obtained with 431~\invfb of
data collected at the $\FourS$ resonance with the \babar\ detector
at the \pep2\ \epem collider at SLAC, corresponding to $474\times 10^{6}$
$\BB$ pairs. An additional ``off-resonance'' data sample of
45~\invfb, collected at a center-of-mass (CM) energy 40~\mev below the
$\FourS$ resonance, is used to study backgrounds from ``continuum''
events, $e^+ e^- \to q \bar{q}$ ($q=u,d,s,$ or $c$).

\section{Event Reconstruction and Selection}

The \babar\ detector is described in detail
elsewhere~\cite{cite:det}. Charged-particle tracking is performed by
a five-layer silicon vertex tracker (SVT) and a 40-layer drift
chamber (DCH). In addition to providing precise position information
for tracking, the SVT and DCH measure the specific ionization, which
is used for identification of low-momentum charged
particles. At higher momenta pions and kaons are
distinguished by Cherenkov radiation detected in a ring-imaging device
(DIRC). The positions and energies of photons are measured with an
electromagnetic calorimeter (EMC) consisting of 6580 thallium-doped
CsI crystals. These systems are mounted inside a 1.5~T solenoidal
superconducting magnet. Muons are identified by the instrumented
flux return, which is located outside the magnet.

The event selection is based on studies of off-resonance data
and Monte Carlo (MC) simulations of continuum and $e^+e^-\to
\Upsilon(4S)\to \BB$ events. The \babar\ detector response
is modeled with \textsc{Geant4}~\cite{cite:geant4}. 
We also use {\tt EvtGen}~\cite{cite:EVTGEN} to model 
the kinematics of \B meson decays and {\tt JetSet}~\cite{cite:jetset} 
to model continuum background processes. All selection criteria are
optimized by maximizing the $S/\sqrt{S+B}$ ratio, 
where $S$ and $B$ are the expected numbers of
the opposite-sign signal and background events, respectively. 
In the optimization we assume an opposite-sign branching fraction
of $4\times10^{-6}$~\cite{cite:PDG}.  

The charged kaon and pion identification criteria are based on a likelihood technique.
These criteria are typically 85\% efficient, depending on the momentum and
polar angle, with misidentification rates at the 2\% level. 
The $\piz$ candidates are reconstructed from pairs of photon
candidates with an invariant mass in the interval
$[119,146]~\mevcc$ and with total energy
greater than 200\mev. Each photon should have energy greater than 70~\mev. 

The neutral $D$ meson candidates are reconstructed from a charged
kaon, a charged pion, and a neutral pion. The correlation between
the tails in the distribution of the $K\pi\piz$ invariant mass,
$m_{D}$, and the \piz candidate mass, $m_{\piz}$, is taken into
account by requiring $|m_{D}-m_{\piz}|$ to be within $24~\mevcc$
of its nominal value~\cite{cite:PDG}, which is $1.5$ times the experimental resolution.

The \Bp candidates are reconstructed by combining $D$ and \Kp
candidates, and constraining them to originate from a common vertex. The
probability distribution of the cosine of the $B$ polar angle with
respect to the beam axis in the CM frame, $\cos\theta_{B}$, is
expected to be proportional to $(1-\cos^2\theta_{B})$. We require
$|\cos\theta_{B}|<0.8$.

We measure two almost independent kinematic variables: the
beam-energy substituted mass
$\mes\equiv\sqrt{(s/2+\vec{p}_0\cdot\vec{p}_B)^2/E^{2}_{0}-{p_B}^2}$,
and the energy difference $\DeltaE \equiv E_B-\sqrt{s}/2$, where
$E$ and $\vec p$ are the energy and momentum, the subscripts $B$ and $0$
refer to the candidate $B$ meson and $e^+e^-$ system, 
respectively, $\sqrt{s}$ is the center-of-mass energy, and
$E_B$ is measured in the CM frame. For correctly reconstructed \B mesons the distribution of 
$\mes$ peaks at the $B$ mass, and the distribution of $\DeltaE$ peaks at zero. The $B$ candidates are
required to have $\DeltaE$ in the range $[-23,23]$~\mev ($\pm 
1.3$ standard deviations). 
We consider only events with \mes\ in the range
$[5.20,5.29]$~\gevcc. 

In less than 2\% of the events, multiple \Bp candidates are present, and in these cases we choose 
that with a
reconstructed $D$ mass closest to the nominal mass
value~\cite{cite:PDG}. If more than one \Bp candidate share the
same $D$ candidate, we select that  
with the smallest $|\DeltaE|$. In the following we refer to the selected 
candidate as $B_{\rm sig}$.
All charged and neutral reconstructed particles not associated with
$B_{\rm sig}$, but with the other \B\ decay in the event, $B_{\rm other}$, are called the rest of the event.

\section{Background Characterization}

After applying the selection criteria described above, the remaining
background is composed of non-signal \BB events and continuum
events. Continuum background events, in contrast to \BB events, are
characterized by a jet-like topology. This difference 
can be exploited to discriminate between 
the two categories of events by means of a Fisher
discriminant \fish, 
which is a linear combination of six
variables. The coefficients of the linear combination are 
chosen to maximize the separation between
signal and continuum background so 
that \fish\ peaks at $1$ for signal and at $-1$ for
continuum background. They are determined with samples of
simulated signal and continuum events, and validated using
off-resonance data. 
In the Fisher discriminant we use the absolute value of the cosine of
the angle between $B_{\rm sig}$ and 
$B_{\rm other}$ thrust axes,
where the thrust axis is defined as the direction
maximizing the sum of the longitudinal momenta of all the
particles. Other variables included in $\fish$ are the event shape
moments $L_0=\sum_{i} p_i$, and $L_2 =\sum_{i} p_i |\cos
\theta_i|^2$, where the index $i$ runs over all tracks and energy
deposits in the rest of the event; $p_i$ is the momentum; and
$\theta_i$ is the angle with respect to the thrust axis of the $B_{\rm sig}$. 
These three variables are calculated in the CM system. We also
use the distance between the decay vertices
of $B_{\rm sig}$ and $D$, the distance of closest approach
between $K$ meson tracks belonging to signal decay chain, and $|\deltat|$, the absolute value of the
proper time interval between the $B_{\rm sig}$ and $B_{\rm other}$
decays~\cite{Aubert:2002rg}. The latter 
is calculated using the measured separation
along the beam direction between the decay points of 
$B_{\rm sig}$ and $B_{\rm other}$ and the Lorentz boost of
the CM frame.  The $B_{\rm other}$ decay point is obtained from 
tracks that do not belong to the reconstructed $B_{\rm sig}$, with constraints
from the $B_{\rm sig}$ momentum and the beam-spot location.
We use \mes and {\fish} to define two regions: the fit region,
defined as $5.20<\mes<5.29 \gevcc$ and $-5< \fish <$~$5$, and the
signal region, defined as $5.27<\mes<5.29~\gevcc$ and $0< {\fish} <
5$.

The \BB background is divided into two components: non-peaking
(combinatorial) and peaking. The latter consists of \B-meson
decays that have a well-pronounced peak in the {\mes} signal region.
One of the decay channels which can mimic opposite-sign signal
events, is the $\Bp\to D\rho^+$ decay with $D\to\Kp\Km$ and
$\rho^+\to\pi^+\piz$. In order to reduce this contribution, we
veto events for which the invariant $\Kp\Km$ pair mass $m_{\Kp\Km}$ is 
$|m_{\Kp\Km}-M_{D{\rm (PDG)}}|>20~\mevcc$ (with the $D$ meson invariant mass, $M_{D{\rm (PDG)}}$,
taken to be $1864.83~\mevcc$~\cite{cite:PDG}). Simulations indicate that the 
remaining background is negligible.

Another possible source of peaking \BB background is the decay
$\Bp\to D\pip$ with $D\to\Kp\pim\piz$, which can
contribute to the signal region of the same-sign sample due to the
misidentification of the \pip as a \Kp. The number of events is expected
to be about 8\% of the total same-sign signal sample (see
Table~\ref{tab:sample}).

The charmless $\Bp\to \Kp\Km\pip\piz$ decay can also
contribute to the signal region.
The branching fraction of this decay has not been measured. Therefore 
the size of this background is estimated from the sidebands of the reconstructed $D$ mass, 
$1.904<M_{D}<2.000~\gevcc$ or
$1.700<M_{D}<1.824~\gevcc$. The result of the study is reported in 
Table~\ref{tab:sample}. In the final fit, we fix the yield of
the same-sign \BB peaking background to the sum of charmless and
open-charm events. The opposite-sign background  in the final event sample is assumed to be negligible. 

The overall reconstruction efficiency for signal events is $(9.6
\pm 0.1)\%$ for opposite-sign signal events and $(9.5 \pm 0.1)\%$ for
same-sign signal events. These numbers are equal within the uncertainty as expected. 
The composition of the final
sample is shown in Table~\ref{tab:sample}.

\begin{table*}
\begin{center}
\caption{\label{tab:sample}Composition of the final selected
sample as evaluated from the MC samples normalized to data and from data 
for the
charmless peaking background. The signal contribution is estimated
using values of branching fractions from the PDG~\cite{cite:PDG} and
$\rB=0.1$~\cite{cite:UTFIT}. The errors are from the statistics of the control samples only.}
\begin{tabular}{lcccccc}
\hline\hline
Sample& Region      & Signal        & \BB non-peaking   & Continuum             & $D\pi$            & Charmless peaking\\
\hline
Same sign       &Fit    & $2252\pm 20$  & $459 \pm 12$      & $7403 \pm 62$     & $176 \pm 14$      &$28\pm 14$\\
                &Signal & $1921\pm 18$  & $147 \pm 8$       & $203 \pm 10$      & $130 \pm 14$      &$21\pm 14$\\
Opposite sign   &Fit    & $28.7\pm 0.2$ & $434 \pm 12$      & $21201 \pm 104$   & -                 &$-2\pm 9$\\
                &Signal & $24.4\pm 0.2$ & $65 \pm 5$        & $612 \pm 18$      & -                 &$-2\pm 9$\\
\hline\hline
\end{tabular}
\end{center}
\end{table*}

\section{Fit Procedure and Results}
To measure the ratios \rplus\ and \rminus\ we perform extended maximum-likelihood fits to the \mes\ and \fish\
distributions, separately for the \Bp\ and \Bm\ data samples. We write
the extended likelihood functions $\mathcal{L}^{\pm}$ as:
\begin{eqnarray*}
&\mathcal{L}^{\pm} = \frac{e^{- N'}}{N !}\cdot {N'}^{N} \cdot \prod_{j=1}^{N} f^{\pm}
({\bf x}_{j} \mid {\bf \theta}, N')\; , \\
& \mbox{with}\;\;f^{\pm}({\bf x} \mid {\bf \theta}, N')=
\frac{1}{N'}\left(\frac{R^{\pm} N_{B^{\pm},{\rm total}}}{1+R^{\pm}}f^{\pm}_{\rm sig,os}({\bf x}| 
{\bf \theta_{\rm sig,os}})+\right. \\
&\left.\frac{N^{\phantom{\pm}}_{B^{\pm},{\rm total}}}{1+R^{\pm}}f^{\pm}_{\rm sig,ss}({\bf x}|
{\bf \theta_{\rm sig,ss}})+\sum_{i} N^{\rm bkg}_{B_{i}} f^{\pm}_{B_{i}}({\bf
x}|{\bf \theta})\right);
\end{eqnarray*}
where $f_{\rm sig, ss}({\bf x}| {\bf \theta_{\rm sig,ss}})$, $f_{\rm
sig, os}({\bf x}| {\bf \theta_{\rm sig,os}})$, and $f_{B_{i}}({\bf
x}|{\bf \theta})$ are the probability density functions (PDFs) of
the hypotheses that the event is a same-sign signal, opposite-sign signal, or a
background event ($B_{i}$ are the different background categories
used in the fit), respectively, $N$ is the number of events in the
selected sample, and $N'$ is the expectation value for the total number
of events. The symbol
\begin{math}{\bf \theta}\end{math} indicates the set of parameters to be fitted.
$N_{B^{\pm}, {\rm total}}$ is the total number of signal events,
$R^{\pm}=\frac{N_{\rm sig, os}}{N_{\rm sig, ss}}$ for the decays of
the $B^{\pm}$ meson, and $N^{\rm bkg}_{B_i}$ is the total number of events of each background
component. For the opposite-sign events the background comes
from continuum and \BBbar\ events. The peaking \BB background 
is introduced as a separate component in the fit to the same-sign sample. The fit is performed to the 
\Bp sample (consisting of 15706 events) to determine \rplus\ and to the \Bm sample (consisting of 15057 events) to determine 
\rminus. The PDFs for \rplus\ and \rminus\ fits are identical.
The \rads\ ratio is fitted to the same likelihood ansatz, but to the combined \Bp\ and \Bm data sample.

Since the correlations among the
variables are negligible, we write the PDFs as products of the
one-dimensional distributions of \mes\ and \fish. The absence of
correlation between these distributions is checked using MC
samples. The signal \mes\ distributions is modeled with the same
asymmetric Gaussian function for both same-sign and opposite-sign
events, while the \fish\ distribution is taken as a sum of two
Gaussians. The continuum background \mes\ distributions for the same and
opposite-sign events are modeled
with two different threshold ARGUS functions~\cite{cite:argus} 
defined as follows:
\begin{equation}
A(x) = x\sqrt{1-\left(\frac{x}{x_0}\right)^2}\cdot
e^{c\left(1-\left(\frac{x}{x_0}\right)^2\right)},\label{eq:threshold}
\end{equation}
where $x_0$ represents the maximum allowed value for the variable
$x$, and $c$ determines the shape of the distribution. The \mes\
distribution of the non-peaking \BB background components are modeled
with Crystal Ball (CB) functions that are different for same-sign and
opposite-sign events~\cite{cite:CB}. The CB function is a Gaussian
modified to include a power-law tail on the low side of the peak.
The \fish\ distributions for the \BB background are approximated with 
sums of two asymmetric Gaussians. For the peaking \BB background we
conservatively use the same parameter set as for the signal.

The PDF parameters are derived from data when possible. The
parameters for continuum events are determined from the
off-resonance data sample. The parameters for the \mes\ distribution
of signal events are extracted from the sample of $\Bp\to D\pip$
with $D\to\Kp\pim\piz$, while for the parameters of the 
signal Fisher PDF we use the MC sample. The parameters of non-peaking
\BB distributions are
determined from the MC sample. 

From each fit, we extract the ratios \rplus, \rminus, or \rads, the total number of
signal events in the sample ($N_{\Bpm, {\rm tot}}$) along with the non-peaking background yields and threshold function slope
for the continuum background. We fix the number of peaking \BB background events.

To test the fitting procedure we generated 10000 pseudo-experiments based on the PDFs described above.
The fitting procedure is then tested on
these samples. We find no bias in the number of fitted events for
any component of the fit. Tests of the fit procedure performed on
the full MC samples give values for the yields compatible with those expected. 

\begin{table}[htpb]
\caption{Results of fits to the \Bp, \Bm, and the combined \Bp and \Bm samples, including 
the extracted number of signal and background events and their statistical errors.\label{tab:fitres}}
\begin{center}
\begin{tabular}{c c c c}
\hline\hline
Sample            & $\Bp$          &  $\Bm$    & $\Bp$ and $\Bm$       \\
\hline
$R,\,10^{-3}$       & $5^{+12}_{-10} $     & $12^{+12}_{-10} $&$9.1^{+8.2}_{-7.6}$\\
$N_{\Bpm, {\rm tot}}$       & $1032\pm 41$      & $946\pm 39$   &$1981\pm 57$\\
$N^{\rm bkg}_{\BB, {\rm OS}}$    & $305\pm 52$       & $120\pm 36$   &$402\pm 65$\\
$N^{\rm bkg}_{\BB, {\rm SS}}$    & $315\pm 44$       & $329\pm 44$   &$644\pm 62$\\
$N^{\rm bkg}_{\rm cont, OS}$  & $10290\pm 111$    & $10017\pm 105$&$20329\pm 154$\\
$N^{\rm bkg}_{\rm cont, SS}$  & $3660\pm 69$      & $3539\pm 68$   &$7203\pm 76$\\
\hline\hline
\end{tabular}
\end{center}
\end{table}

The main results of the fit to the data
are summarized in Table~\ref{tab:fitres}.

\begin{figure*}[t!]
\begin{center}
\epsfig{file=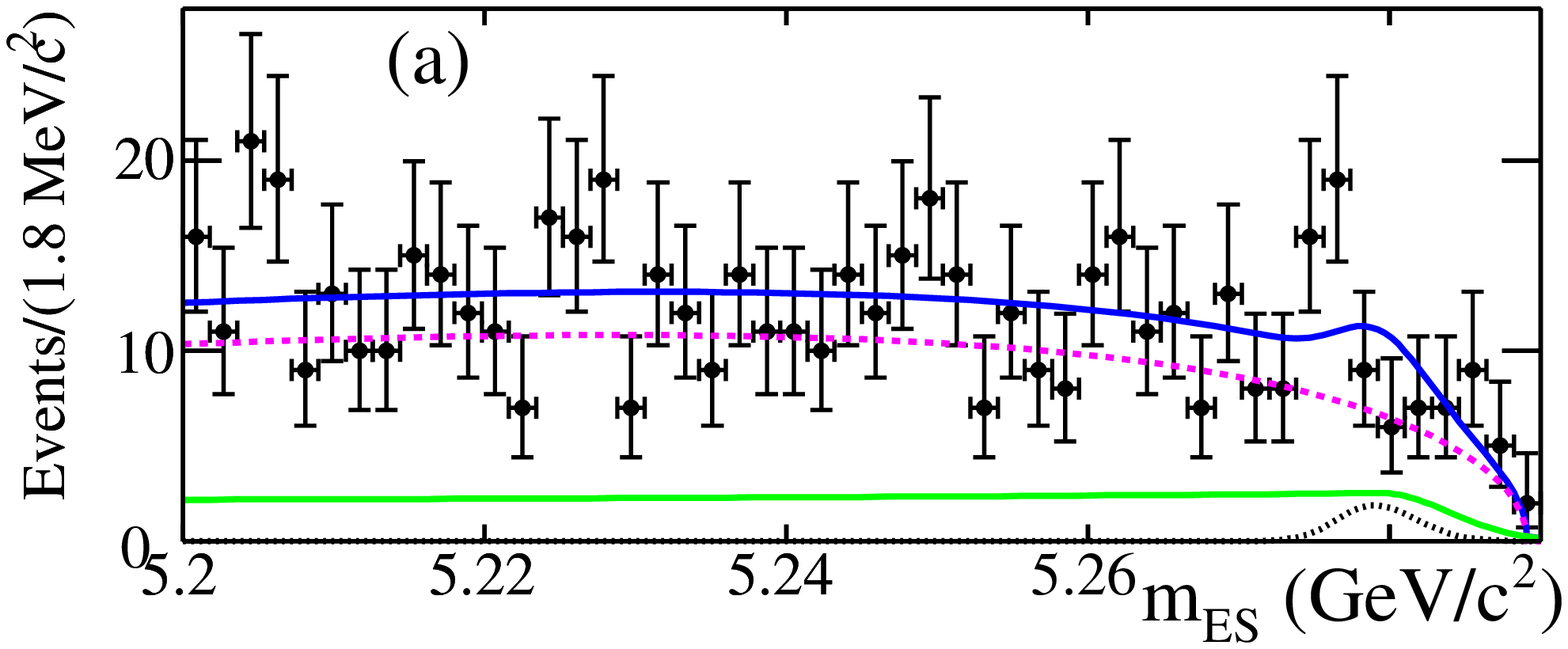,width=0.4\linewidth}
\epsfig{file=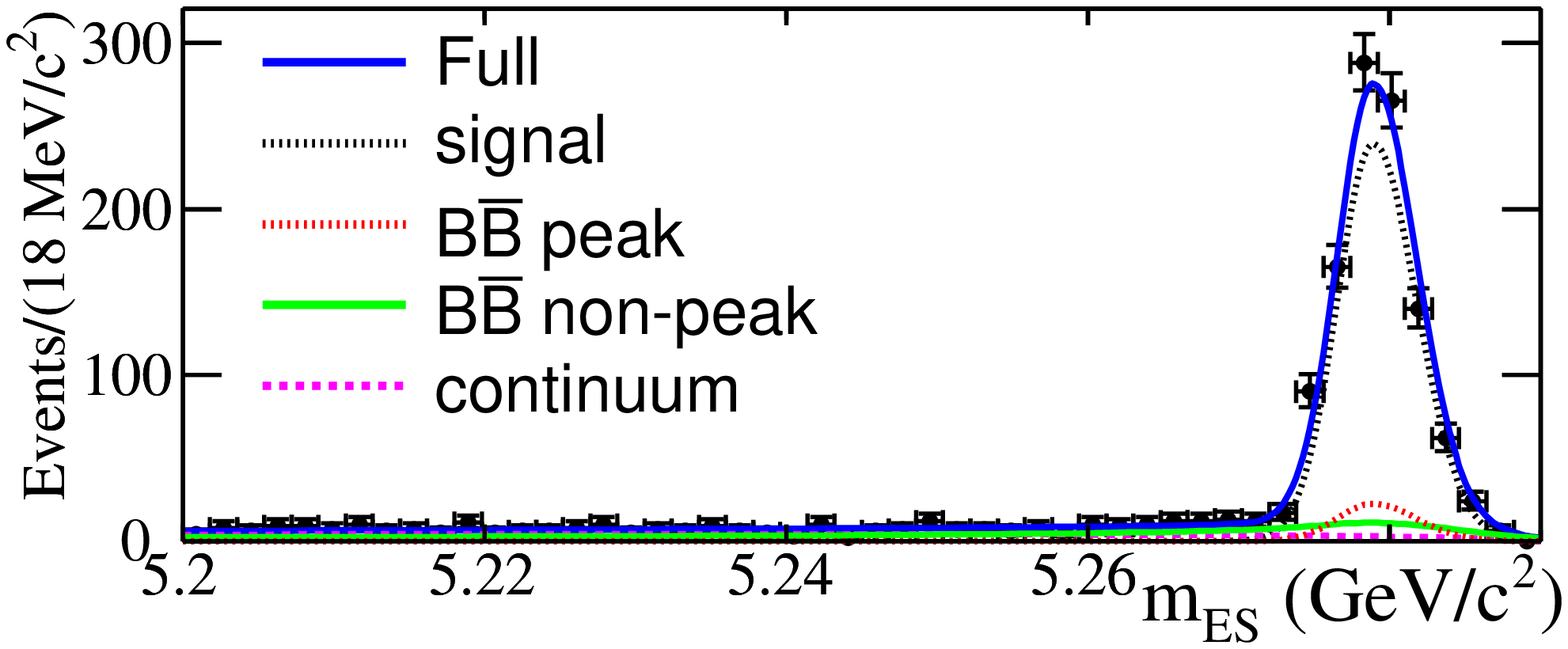,width=0.4\linewidth}
\epsfig{file=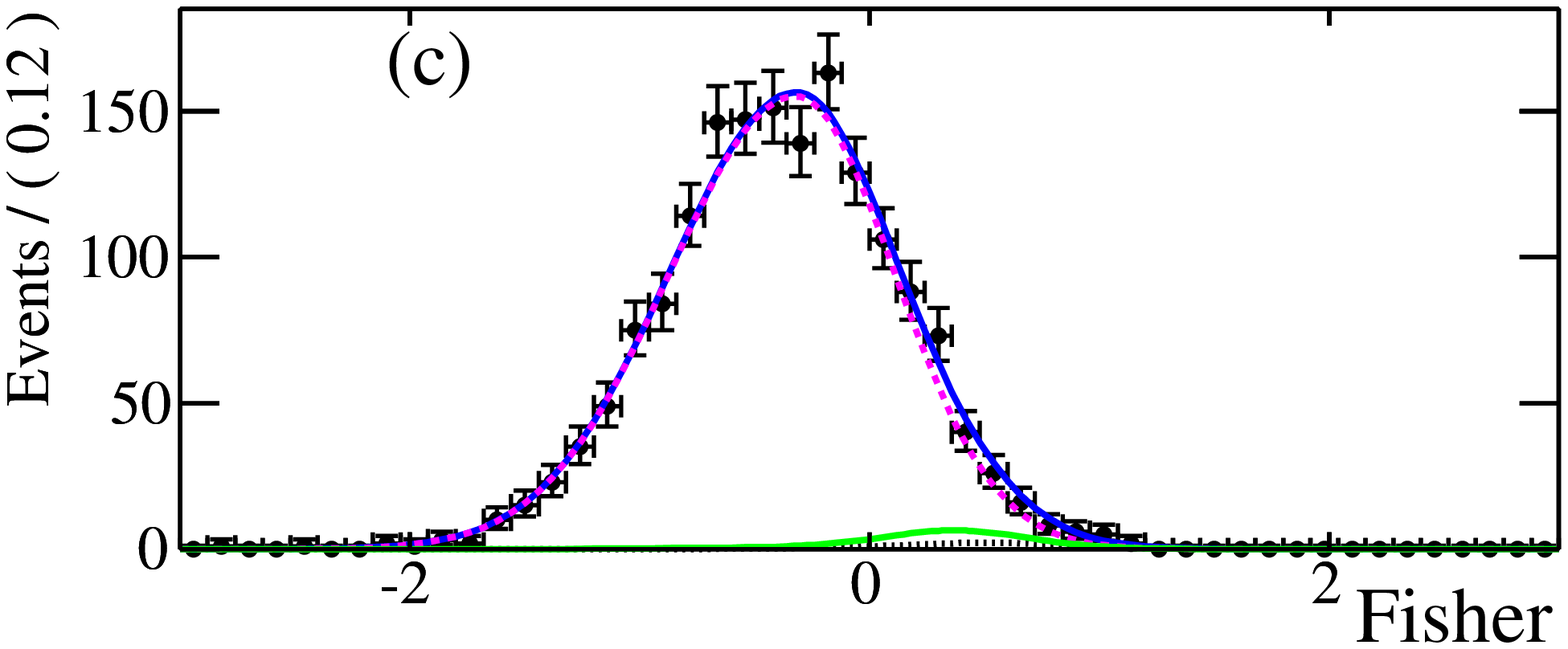,width=0.4\linewidth}
\epsfig{file=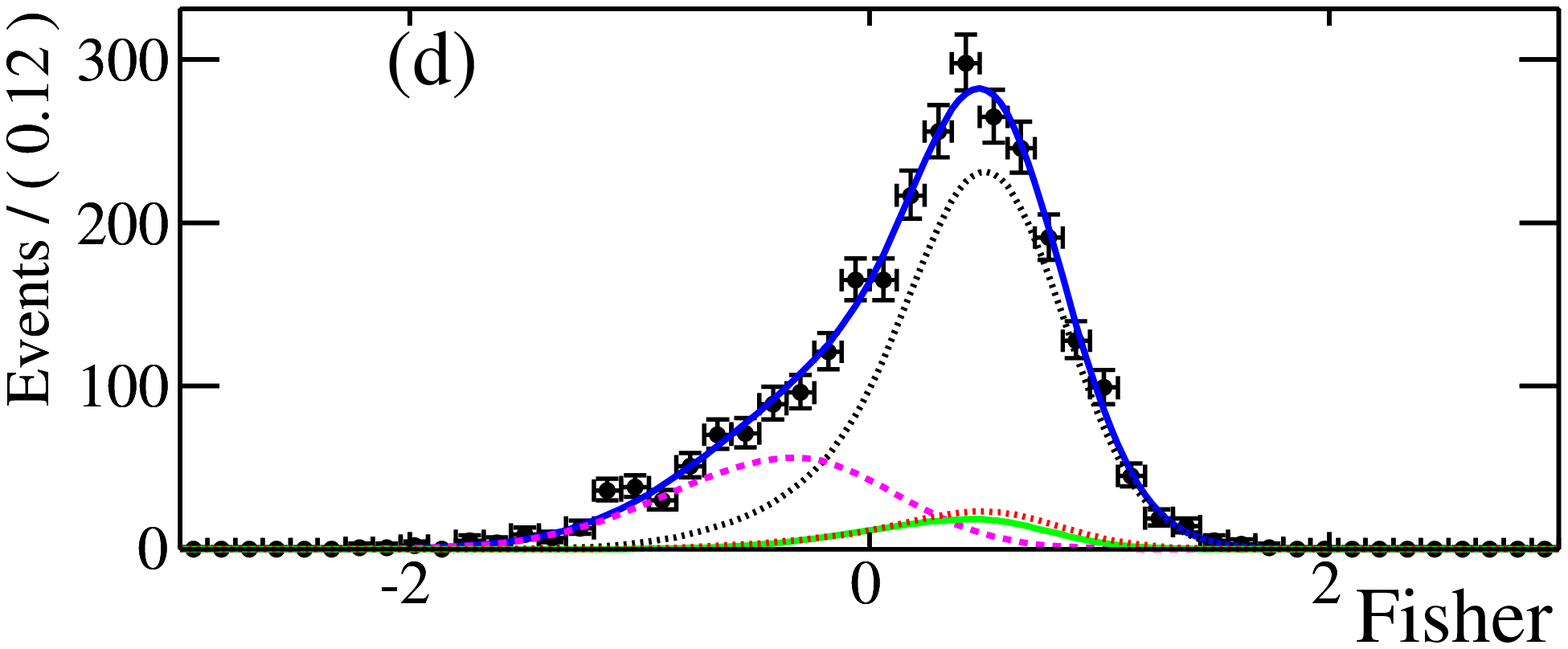,width=0.4\linewidth}
\end{center}
\caption{(color online) Distribution of (a,b) \mes\ (with ${\fish} >0.5$) and (c,d) {\fish} (with $\mes>5.27~\gevcc$) 
and the results of the maximum likelihood fits for the combined \Bp and \Bm samples (extracting \rads),
for (a,c) opposite-sign and (b,d) same-sign decays. The data are well described by the overall fit result (solid blue line) 
which is the sum of the signal, continuum, non-peaking, and peaking \BB backgrounds.  
\label{fig:Radsfit}}
\end{figure*}

\begin{figure*}[t!]
\begin{center}
\epsfig{file=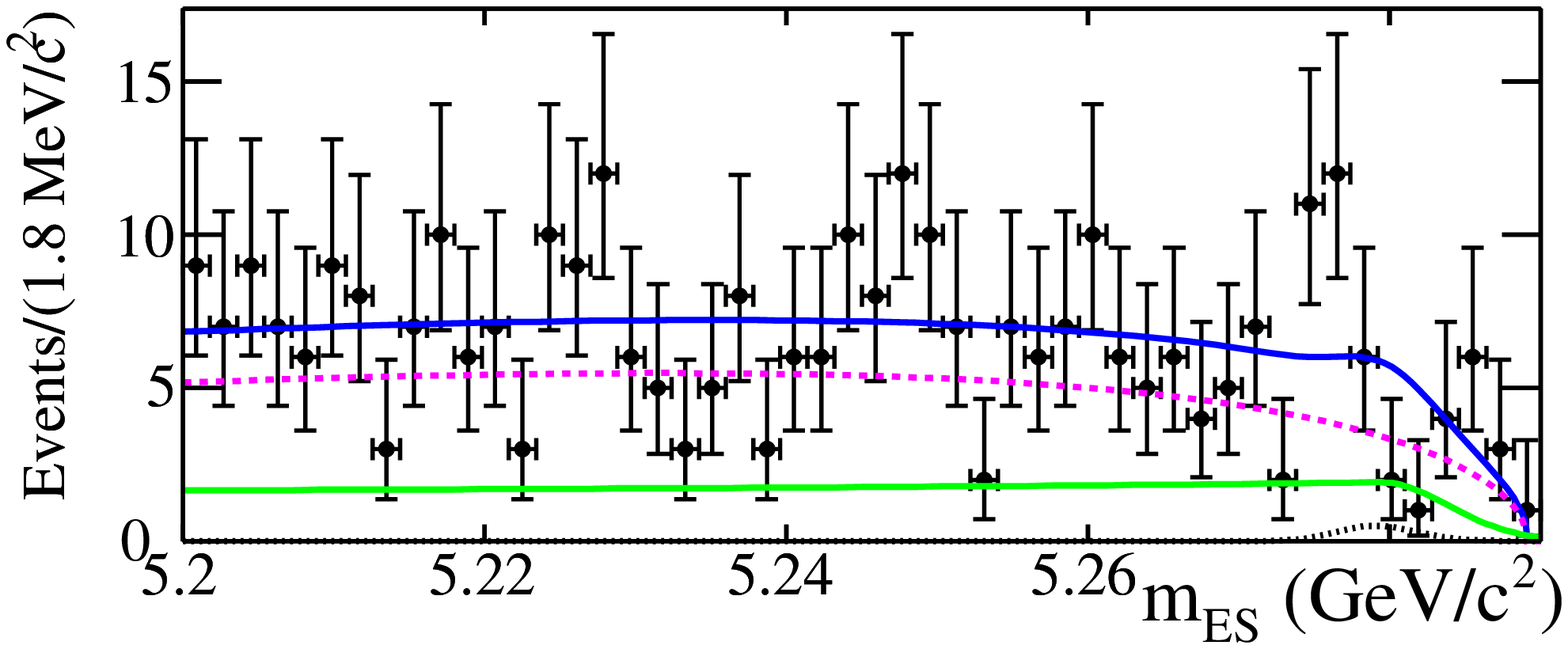,width=0.4\linewidth}
\epsfig{file=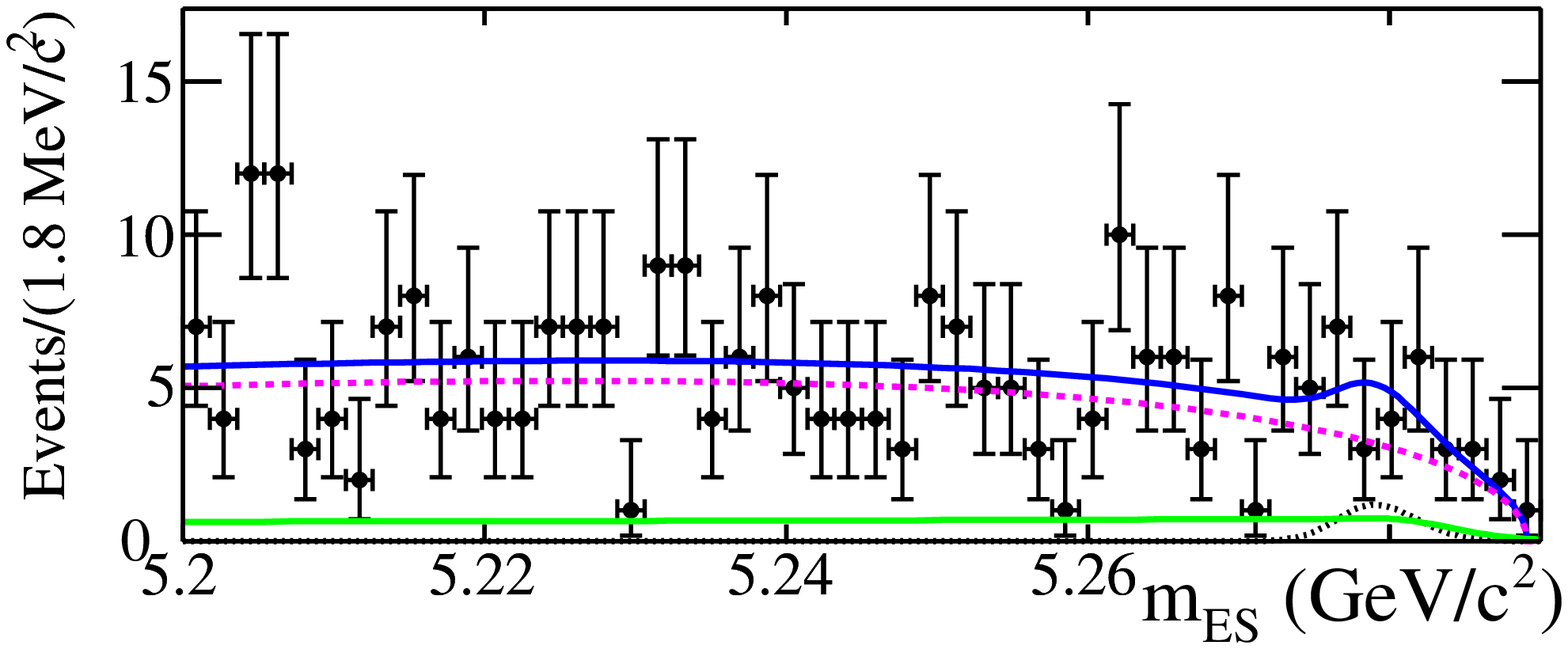,width=0.4\linewidth}
\end{center}
\caption{(color online) Projections of the 2D likelihood for \mes with
the additional requirement 
${\fish}>0.5$, obtained from the fit to the \Bp (left) and
\Bm (right) data sample for opposite-sign events (extracting \rplus\ and \rminus). 
The labeling of the curves is the same as in Fig.~\ref{fig:Radsfit}.\label{fig:Rplusfit}}
\end{figure*}

The fits to the \mes\ for ${\fish}>0.5$ and the \fish\ distribution with ${\mes}>5.27~\gevcc$ 
are shown in Fig.~\ref{fig:Radsfit}, for the combined \Bp\ and \Bm\ sample. 
These restrictions reduce the background and retain most of the signal events. 
Fig.~\ref{fig:Rplusfit} shows the fits for the separate \Bp\ and \Bm\ samples.

\section{Systematic Uncertainties}

We consider various sources of systematic uncertainties, listed in Table~\ref{tab:systMC_ADS}. 
One of the largest
contributions comes from the uncertainties on the PDF
parameters. To evaluate the contributions related to the
$m_{\rm ES}$ and \fish\ PDFs, we repeat the fit varying the
PDF parameters for each fit species within their statistical errors, taking
into account correlations among the parameters (labeled as ``PDF
error'' in Table~\ref{tab:systMC_ADS}).

To evaluate the uncertainties arising from peaking background
contributions, we repeat the fit varying the the peaking \BB background contribution within its statistical uncertainties
and the errors of branching 
fractions, \BR, used
to estimate the contribution. For the 
opposite-sign events only the positive part of the probability distribution
is used in the evaluation.

Differences between data and MC (labeled as ``Simulation'' in Table~\ref{tab:systMC_ADS}) in
the shape of the \fish\ distribution are studied for signal components
using the data control samples of $\Bp \to D \pip$ with
$D\to\Kp\pim\piz$. These parameters are expected to be slightly different
between the $\B\to D\pi$ and $\B\to DK$ samples.
We conservatively take the systematic uncertainty
as the difference in the fit results from the nominal
parameters set (using MC events) and the parameters set obtained using
the $\B\to D\pi$ data sample.

The systematic uncertainty attributed to the crossfeed between
opposite-sign and same-sign events has been evaluated from the MC
samples. The number of same-sign events passing the selection of the
opposite-sign events is taken as a systematic uncertainty. 
The efficiencies for same-sign and
opposite-sign events were verified to be the same within a precision of
3\%~\cite{cite:ViolaB0}. We
hence assign a systematic uncertainty on $R^{\pm}$ based on variations due to changes in the efficiency ratio by $\pm 3\%$.

The systematic uncertainties for the ratios \rplus, \rminus, and \rads\ are summarized in Table~\ref{tab:systMC_ADS}.
The overall systematic errors represent the sum in quadrature of the individual uncertainties.

\begin{table}
\begin{center}
\caption{Systematic errors for
$R^{\pm}$ and \rads\ in units of $10^{-3}$.\label{tab:systMC_ADS}}
\begin{tabular}{lccc}
\hline\hline
Source&\rplus\ &\rminus\ &\rads\ \\
\hline
PDF error     &$^{+1.1}_{-1.8}$&1.1&1.0\\
Same sign peaking background &0.2&0.5&0.2\\
Opposite sign peaking background &$^{+0}_{-3.6}$&$^{+0}_{-3.6}$&$^{+0}_{-3.4}$\\
Simulation  &0.6&0.6&0.7\\
\BR\ errors     &0.2&0.6&0.4\\
Crossfeed contribution &0.1&0.4&0.3\\
Efficiency ratio  &0.1&0.4&0.3\\
\hline
Combined uncertainty&$^{+1.2}_{-4.1}$&$^{+1.6}_{-3.9}$&$^{+1.4}_{-3.7}$\\
\hline\hline
\end{tabular}
\end{center}
\end{table}

\section{Extraction of $\mbox{\text{\boldmath$\rB$}}$}

\begin{figure}[h]
\begin{center}
\epsfig{file=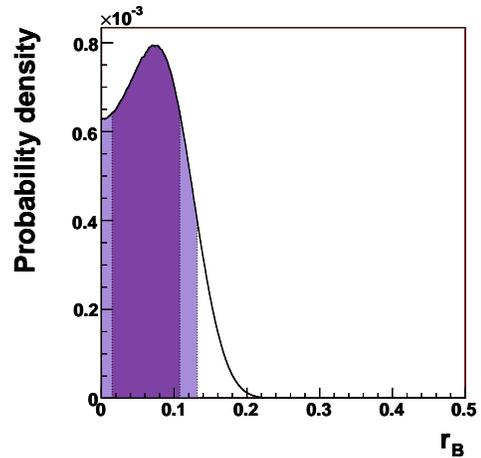,width=0.8\linewidth}
\end{center}
\caption{Bayesian posterior probability density 
function for \rB\ from our measurement of
\rplus\ and \rminus\ and the hadronic $D$ decay 
parameters \rD, \deltaD, and \kD\ taken 
from~\cite{cite:CLEO} and~\cite{cite:PDG}. 
The dark and light shaded zones represent the
68\% and 90\% probability regions, respectively. \label{fig:rB}}
\end{figure}

Following a Bayesian approach~\cite{cite:bayes}, the probability distributions for the 
$\rplus$ and $\rminus$ ratios obtained in the fit
are translated into a probability distribution for $\rB$ using 
Eqs.~\ref{eq:main1} and~\ref{eq:main2} simultaneously. We assume the following prior 
probability distributions: for \rD\ a Gaussian with mean 
$4.7\times 10^{-2}$ and standard deviation $3\times 10^{-3}$~\cite{cite:PDG}; 
for \kD\ and \deltaD, we use the likelihood obtained in Ref.~\cite{cite:CLEO}, 
taking into account a 180 degree difference in the phase convention for \deltaD;
 for \g\ and \deltaB\ we assume a uniform distribution 
between 0 and 360 degrees, while for \rB\ a uniform 
distribution in the range $[0,1]$ is used.
We obtain the posterior probability distribution shown in Fig.~\ref{fig:rB}.
Since the measurements are not statistically significant, we
integrate over the positive portion of that distribution 
and obtain the upper limit
$\rB\ < 0.13$ at 90\% probability,
and the range
\begin{equation}
\rB \in [0.01,0.11]\text{ at 68\% probability,}
\end{equation}
and 0.078 as the most probable value.

\section{Summary}

We have presented a study of the decays $\Bpm \to \Dz
\Kpm$ and $\Bpm \rightarrow \Dzb \Kpm$, in which the $\Dz$ and $\Dzb$ mesons 
decay to the $\Kmp\pipm\pi^0$ final state 
using the ADS method. The analysis is performed using
$474\times10^{6}$~\BB pairs, the full \babar\ dataset. Previous
results~\cite{cite:ViolaADS} are improved and superseded 
by improved event reconstruction algorithms and analysis strategies 
employed on a larger data sample.

The final results are:
\begin{eqnarray}
\rplus\ = \left(5^{+12}_{-10}\stat
^{+1}_{-4}\syst\right)\times 10^{-3},\\
\rminus\ = \left(12^{+12}_{-10}\stat
^{+2}_{-4}\syst\right)\times 10^{-3},\\
\rads\ = \left(9.1^{+8.2}_{-7.6}\stat
^{+1.4}_{-3.7}\syst\right)\times 10^{-3},
\end{eqnarray}
from which we obtain 90\% probability limits:
\begin{eqnarray}
\rplus\ < 23 \times 10^{-3},\\
\rminus\ < 29 \times 10^{-3},\\\
\rads\ < 21\times 10^{-3}.
\end{eqnarray}
From our measurements we derive the limit
\begin{equation}
{\rB}<0.13 \text{ at 90\% probability}.
\end{equation}

\section{Acknowledgments}

\input{acknowledgements.tex}

\end{document}

%% file: authors_feb2011.tex
%
\author{J.~P.~Lees}
\author{V.~Poireau}
\author{V.~Tisserand}
\affiliation{Laboratoire d'Annecy-le-Vieux de Physique des Particules (LAPP), Universit\'e de Savoie, CNRS/IN2P3,  F-74941 Annecy-Le-Vieux, France}
\author{J.~Garra~Tico}
\author{E.~Grauges}
\affiliation{Universitat de Barcelona, Facultat de Fisica, Departament ECM, E-08028 Barcelona, Spain }
\author{M.~Martinelli$^{ab}$}
\author{D.~A.~Milanes$^{a}$}
\author{A.~Palano$^{ab}$ }
\author{M.~Pappagallo$^{ab}$ }
\affiliation{INFN Sezione di Bari$^{a}$; Dipartimento di Fisica, Universit\`a di Bari$^{b}$, I-70126 Bari, Italy }
\author{G.~Eigen}
\author{B.~Stugu}
\author{L.~Sun}
\affiliation{University of Bergen, Institute of Physics, N-5007 Bergen, Norway }
% ...
\collaboration{The \babar\ Collaboration}
\noaffiliation

%% file: acknowledgements.tex
We are grateful for the 
extraordinary contributions of our \pep2\ colleagues in
achieving the excellent luminosity and machine conditions
that have made this work possible.
The success of this project also relies critically on the 
expertise and dedication of the computing organizations that 
support \babar.
The collaborating institutions wish to thank 
SLAC for its support and the kind hospitality extended to them. 
This work is supported by the
US Department of Energy
and National Science Foundation, the
Natural Sciences and Engineering Research Council (Canada),
the Commissariat \`a l'Energie Atomique and
Institut National de Physique Nucl\'eaire et de Physique des Particules
(France), the
Bundesministerium f\"ur Bildung und Forschung and
Deutsche Forschungsgemeinschaft
(Germany), the
Istituto Nazionale di Fisica Nucleare (Italy),
the Foundation for Fundamental Research on Matter (The Netherlands),
the Research Council of Norway, the
Ministry of Education and Science of the Russian Federation, 
Ministerio de Ciencia e Innovaci\'on (Spain), and the
Science and Technology Facilities Council (United Kingdom).
Individuals have received support from 
the Marie-Curie IEF program (European Union), the A. P. Sloan Foundation (USA) 
and the Binational Science Foundation (USA-Israel).

%% file: note2308.bbl
\begin{thebibliography}{99}
\bibitem{cite:charge}    Charge conjugate processes are assumed throughout the paper.
\end{thebibliography}
